\documentclass{PoS}
\usepackage{amsmath}
\usepackage{gensymb}
\usepackage{url}

\let\OLDthebibliography\thebibliography
\renewcommand\thebibliography[1]{
  \OLDthebibliography{#1}
  \setlength{\parskip}{0pt}
  \setlength{\itemsep}{0pt plus 0.3ex}
}

\usepackage{lineno}

\newcommand*\patchAmsMathEnvironmentForLineno[1]{%
  \expandafter\let\csname old#1\expandafter\endcsname\csname #1\endcsname
  \expandafter\let\csname oldend#1\expandafter\endcsname\csname end#1\endcsname
  \renewenvironment{#1}%
     {\linenomath\csname old#1\endcsname}%
     {\csname oldend#1\endcsname\endlinenomath}}%
\newcommand*\patchBothAmsMathEnvironmentsForLineno[1]{%
  \patchAmsMathEnvironmentForLineno{#1}%
  \patchAmsMathEnvironmentForLineno{#1*}}%
\AtBeginDocument{%
\patchBothAmsMathEnvironmentsForLineno{equation}%
\patchBothAmsMathEnvironmentsForLineno{align}%
\patchBothAmsMathEnvironmentsForLineno{flalign}%
\patchBothAmsMathEnvironmentsForLineno{alignat}%
\patchBothAmsMathEnvironmentsForLineno{gather}%
\patchBothAmsMathEnvironmentsForLineno{multline}%
}

\title{Gamma Emission from Large Galactic Structures}

\ShortTitle{Gamma Emission from Large Galactic Structures}

\author{\speaker{Henrike Fleischhack}\\
        Michigan Technological University\\
        E-mail: \email{hfleisch@mtu.edu}}

\author{Hugo Alberto Ayala Solares\thanks{Corresponding author}\\
        The Pennsylvania State University\\
        E-mail: \email{hgayala@psu.edu}}
        
 \author{Petra Huentemeyer\\
 		 Michigan Technological University\\
        E-mail: \email{petra@mtu.edu}}
        
\author{Matthew Coel\\
		 Michigan Technological University\\
        E-mail: \email{mtcoel@mtu.edu}}
        
 \author{for the HAWC Collaboration\thanks{ For a complete author list, see {http://www.hawc-observatory.org/collaboration/icrc2017.php} }}

\abstract{Gamma-ray emission from large structures is useful for tracing the propagation and distribution of cosmic rays throughout our Galaxy. For example, the search for gamma-ray emission from Giant Molecular Clouds may allow us to probe the flux of cosmic rays in distant galactic regions and to compare it with the flux measured at Earth. Also, the composition of the cosmic rays can be measured  by separating the gamma-ray emission from hadronic or leptonic processes. In the case of emission from the Fermi Bubbles specifically, constraining the mechanism of gamma-ray production can point to their origin. HAWC possesses a large field of view and good sensitivity to spatially extended sources, which currently makes it the best suited ground-based observatory to detect extended regions. We will present preliminary results on the search of gamma-ray emission from Molecular Clouds, as well as upper limits on the differential flux from the Fermi Bubbles.}

\FullConference{35th International Cosmic Ray Conference --- ICRC2017\\
		10--20 July, 2017\\
		Bexco, Busan, Korea}

\begin{document}

\section{Introduction}

One of the topics in the modern field of high-energy astrophysics is the origin and propagation of cosmic rays. However, the directional information of the cosmic rays is lost due to their interaction with the interstellar magnetic field. 
The observation of the gamma-ray emission by the interaction of cosmic rays with interstellar media (i.e. radiation fields, matter, etc) is one of the tools that helps trace the origin, acceleration, propagation and distribution of cosmic rays through the galaxy. 

Probing the flux of cosmic rays in distant galactic regions can be achieved by measuring the gamma-ray emission from Giant Molecular Clouds (GMCs) that are located far from cosmic-ray sources, i.e. passive clouds. This is because the gamma-ray emission from GMCs is proportional to the cosmic-ray flux, hence this is an indirect way to measure the galactic cosmic-ray flux and  can be compared to the measured cosmic ray flux at Earth \cite{casanova2010}. 

The gamma-ray signal can also help distinguish between its possible hadronic or leptonic origin, leading to a study of the composition and origin of the  cosmic rays. In the case of emission from the Fermi Bubbles specifically, constraining the mechanism of gamma-ray production can point to their origin \cite{crocker11, Cheng11, Guo12, Mou14} and give an understanding of the evolution of our galaxy.

The HAWC gamma-ray observatory can search for large-scale structures thanks to its large field of view of 2\,sr and high-duty cycle of $> 95\%$. HAWC is sensitive to gamma rays with energies between 100\,GeV and 100\,TeV. It is located on the volcano Sierra Negra in the state of Puebla, Mexico, at an altitude of 4100\,m a.s.l.
HAWC uses the water Cherenkov technique to detect the electromagnetic component of the shower fronts of extensive air showers that reach the ground. From the footprint of the shower, the direction of the primary gamma ray or cosmic ray that interacts with the atmosphere is reconstructed.  In this presentation, data recorded by the HAWC observatory is used to search for gamma-ray signal from large galactic structures.

\section{Giant Molecular Clouds}
GMCs are dense concentrations of interstellar gas containing masses around $10^4 - 10^6 \, M_{\odot}$ and with sizes of $50 - 200 \, \text{pc}$. They are composed mainly of cold, dark dust and molecular  gas --- mostly molecular hydrogen and helium. GMCs are the main factories of stars in the galaxy. 

The gamma-ray flux, produced by the interaction of cosmic rays with the GMC is proportional to
\begin{equation}\label{eq:flux}
F_{\gamma} \propto \Phi_{CR}\frac{M_5}{d_{kpc}^2},
\end{equation}
where $\Phi_{CR}$ is the cosmic-ray flux, $M_5 = M/10^5M_{\odot}$ is the mass of the molecular cloud, $d_{\text{kpc}} = d / 1\,\text{kpc}$ is the distance to the molecular cloud\cite{casanova2010,aharonian90}.
Assuming that $\Phi_{CR}$ is equal to the locally measured cosmic-ray flux, the gamma-ray flux from equation \ref{eq:flux} can be estimated as
\begin{align}\label{eq:flux2}
F_{\gamma}  = \left\{ \begin{array}{cc} 
                1.45\times10^{-13}E_{\text{TeV}}^{-1.75} (M_5/d_{\text{kpc}}^2)  \,\text{cm}^{-2} \, \text{s}^{-1} & \hspace{5mm} 100\,\text{MeV} <E_{\gamma}<\, 1 \, \text{TeV} \\
                 & \\
                2.85\times10^{-13}E_{\text{TeV}}^{-1.6} (M_5/d_{\text{kpc}}^2) \,\text{cm}^{-2} \, \text{s}^{-1} & \hspace{5mm} E_{\gamma}>\, 1 \, \text{TeV} \\
                \end{array}, \right.
\end{align}
where $E_{\text{TeV}} = E/1\,\text{TeV}$ and $F_{\gamma}$ is the energy-integrated flux \cite{aharonian90}.

\begin{figure}
\centering
\includegraphics[scale=0.28]{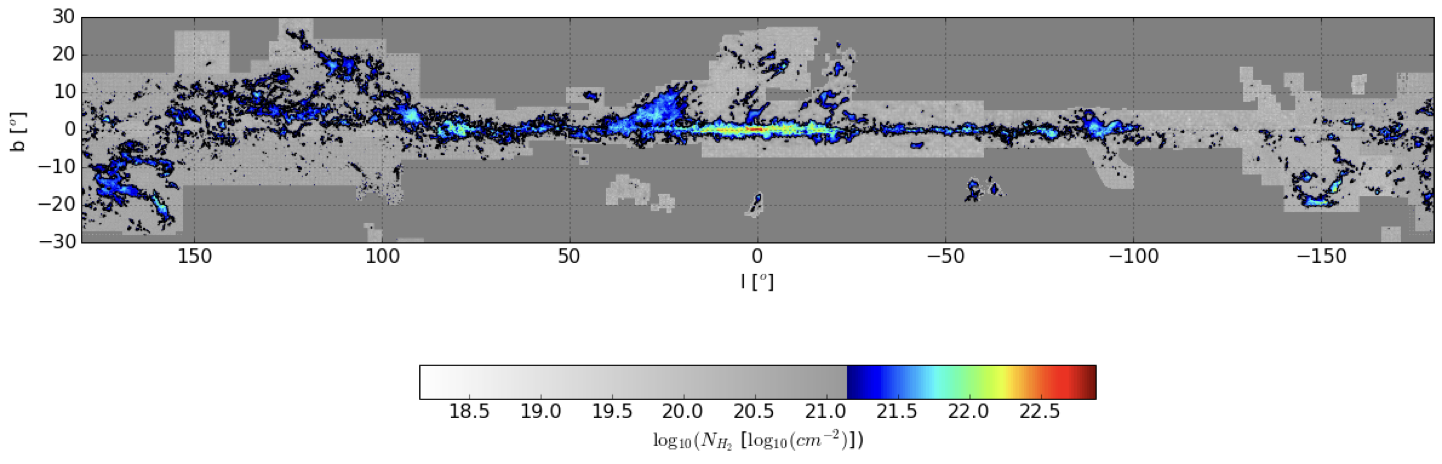}
\caption{Distribution of CO gas from the CfA-Chile survey \cite{dame}.}
\label{fig:survey}
\end{figure}

GMCs have been observed mostly in the radio and infrared part of the electromagnetic spectrum since optical photons are not able to penetrate these dense regions. 
The most recent survey of GMCs has be done by the CfA-Chile survey. The survey is based on the observation of the CO-115\,GHz frequency \cite{dame} and it is shown in Figure \ref{fig:survey}.

\begin{figure}
\centering
\includegraphics[scale=0.4]{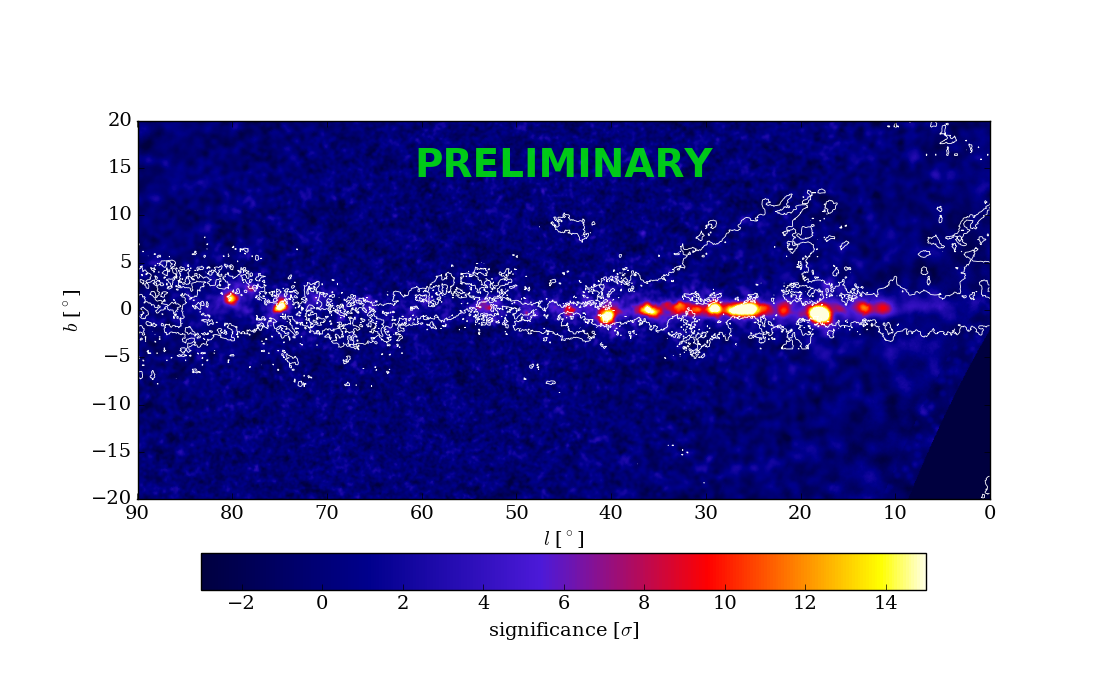}
\caption{HAWC observations with CO maps contour.}
\label{fig:cont}
\end{figure}

An overlap of 760 days of HAWC data, in the galactic region  $0^{\degree} < l < 90^{\degree}$, with a contour from the Cfa-Chile survey is shown in Figure \ref{fig:cont}. 

\subsection{Sensitivity to Molecular clouds}

\begin{figure}
\centering
\includegraphics[scale=0.4]{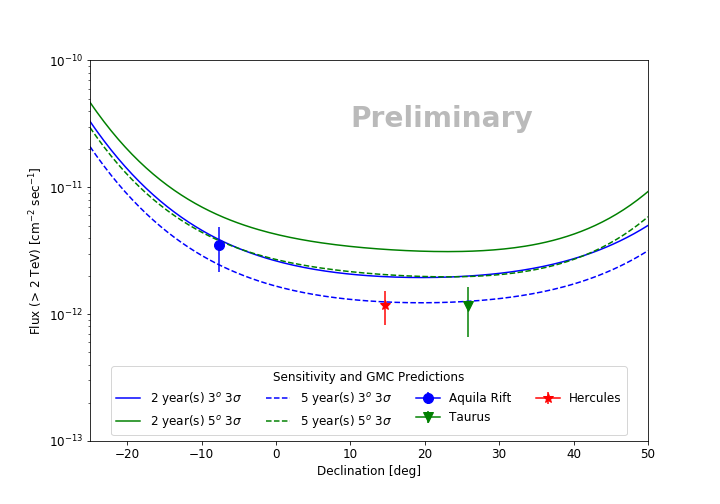}
\includegraphics[scale=0.4]{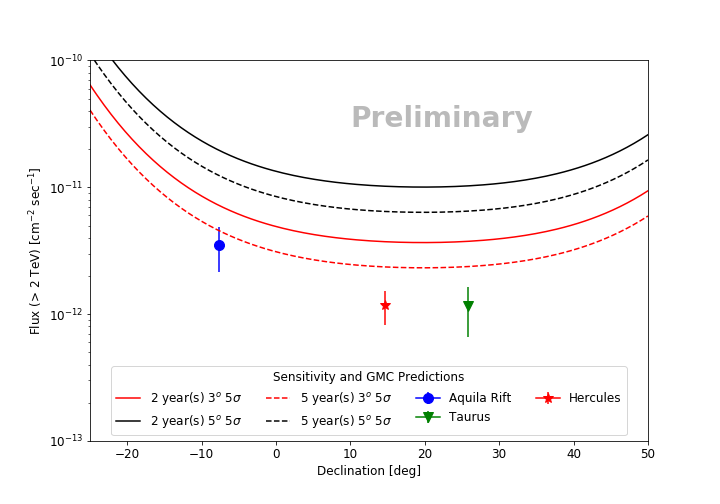}
\caption{HAWC sensitivity to extended sources and predicted integral fluxes of the GMCs in their respective declination. The error bars are calculated from the respective mass and distance errors given by the references.}
\label{fig:sensi}
\end{figure}

Using equation \ref{eq:flux2}, we compare the value of the expected flux to the sensitivity of the HAWC detector to extended sources. For the calculation of the sensitivity we assume disc regions of $3^{\degree}$ and $5^{\degree}$ due to the variety of the morphology of the GMCs.
For the sensitivity we apply the procedure described in \cite{kashyap} \footnote{Named upper limit instead of sensitivity in the publication}. 
The probability for false positives is set to $\alpha = 0.003 \, (3\sigma)$ and $\alpha = 0.000006 \, (5\sigma)$, and the probability of detection is set to $\beta = 0.5$.
Figure \ref{fig:sensi} shows the sensitivity plot compared to the predictions from the GMCs Aquila Rift, Taurus and Hercules. Table \ref{tab:gmcs} describes the properties of the GMCs.
HAWC does not expect to  detect either of these objects with the current dataset (760 days). However, HAWC may be sensitive to the GMCs  at the $3\sigma$ level with its 5 year dataset.

\begin{table}[!h]
\centering
\begin{tabular}{|c|c|c|c|c|}
\hline 
GMC & Mass & Distance & Decl. Center & Extension \\ 
\hline 
Aquila Rift & $1.5\times10^5 \, M_{\odot}$ \cite{aqher} & $225\pm55\,\text{pc}$  \cite{aq1} & $-7.6^{\degree}$ & $<$0.068 sr \\ 
\hline 
Taurus & $0.2\times10^5 \, M_{\odot}$ \cite{tau} & $135 \pm 20\, \text{pc}$ \cite{her1} & $25.8^{\degree}$ &  $<$0.203 sr\\ 
\hline 
Hercules & $0.5\times10^5 \, M_{\odot}$ & $200 \pm 30\,\text{pc}$ \cite{her1} & $14.7^{\degree}$  & $<$0.013sr\\ 
\hline 
\end{tabular} 
\caption{Description of GMCs. The mass of Hercules is assumed since no value was found in the literature.}
\label{tab:gmcs}
\end{table}

%

\begin{figure}[!ht]
\centering
\includegraphics[scale=0.45]{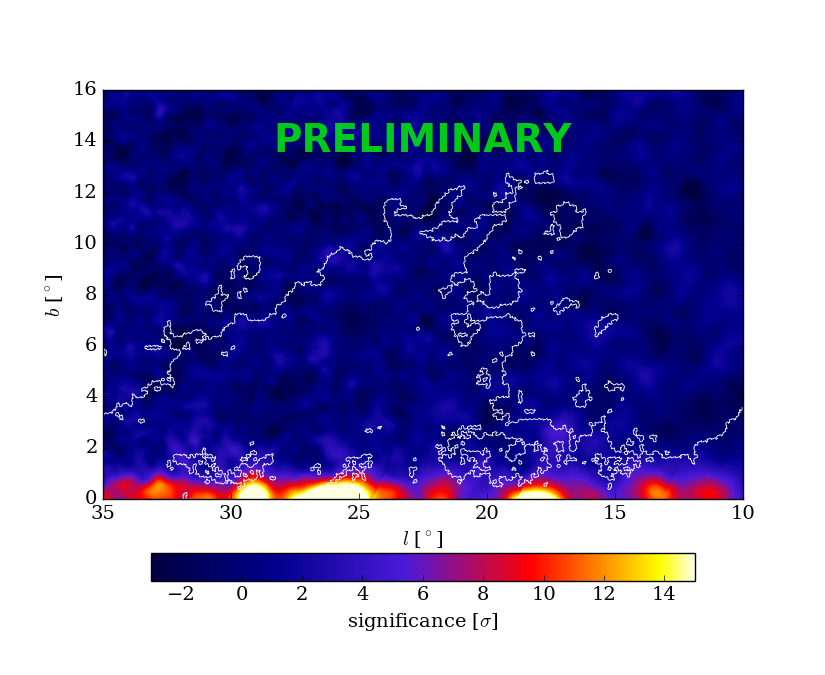}
\caption{Aquila Rift is located in the region $10^{\degree} < l < 35^{\degree}$ and $0^{\degree} < b < 15^{\degree}$. White is the contour line of the CO-gas map. }
\label{fig:aqexc}
\end{figure}

\begin{figure}[!ht]
\centering
\includegraphics[scale=0.45]{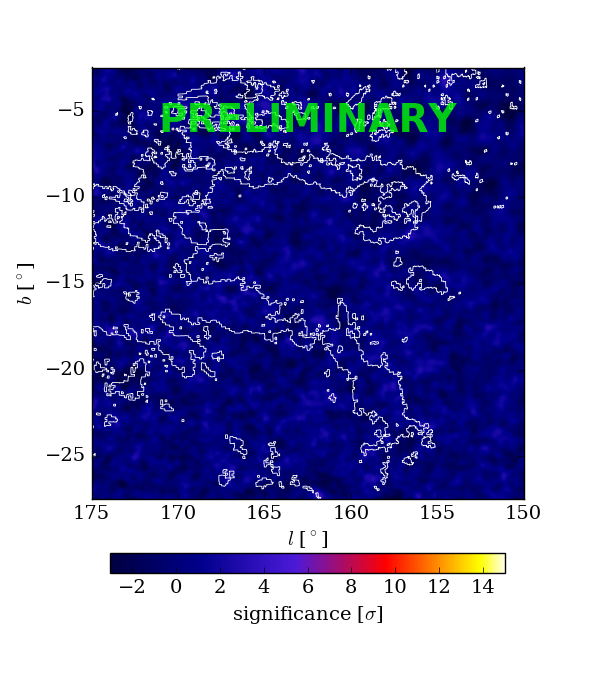}
\caption{Taurus is located in the region  $150^{\degree} < l < 175^{\degree}$ and $-28^{\degree} < b < -2^{\degree}$. White is the contour line of the CO-gas map. }
\label{fig:taexc}
\end{figure}

\begin{figure}[!ht]
\centering
\includegraphics[scale=0.5]{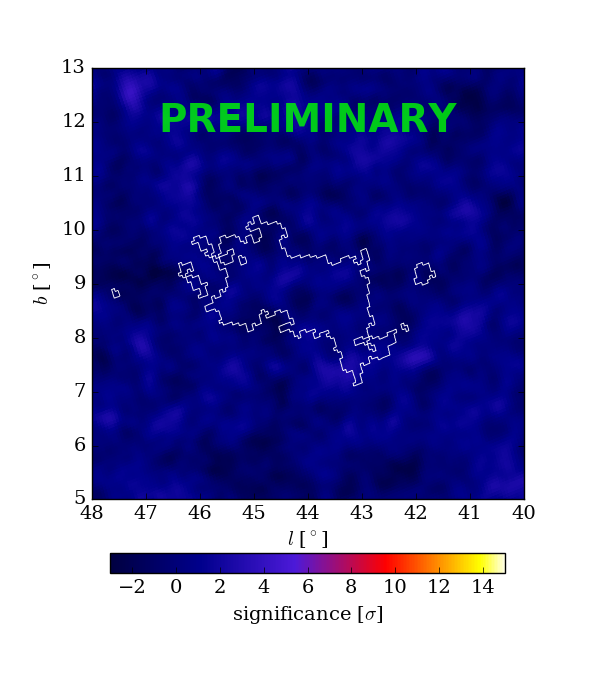}
\caption{Hercules is located in the region  $40^{\degree} < l < 48^{\degree}$ and $7^{\degree} < b < 11^{\degree}$. White is the contour line of the CO-gas map.  }
\label{fig:herexc}
\end{figure}

A zoom-in of the GMCs  in the HAWC map are shown in Figures \ref{fig:aqexc}, \ref{fig:taexc} and \ref{fig:herexc}.

%
%

\section{\textit{Fermi} Bubbles}

The \textit{Fermi}-LAT Bubbles are two bubble-like structures extending $55^{\degree}$ above and below the Galactic plane. They were discovered after looking for a counterpart of the microwave haze in data from the \textit{Fermi} Telescope \cite{bubble1}. 
In the literature exists several models that try to explain the origin of the \textit{Fermi} Bubbles. Some of these ideas include: outflow can be generated by activity of the nucleus in our galaxy producing a jet \cite{Guo12}, wind from long-timescale star formation\cite{crocker11}, periodic star capture processes by the supermassive black hole in the Galactic center \cite{Cheng11} or by winds produced by the hot accretion flow in Sgr A* \cite{Mou14}. 
These models can be constrained by measuring the energy spectrum of the gamma-ray emission. For example, if the gamma-ray emission cannot be explained by hadronic processes, the wind from long-timescale star formation model could be ruled out.

\subsection{Method}
We use HAWC data, corresponding to the dates of 2014 November 27 to 2016 February 11, to search for gamma-ray emission from the Northern \textit{Fermi} Bubble region.
The main challenge of the analysis is to estimate the background. First, we need to distinguish between the shower signatures of cosmic rays and gamma rays that deposit their energy in the HAWC observatory. Then we need to estimate the isotropic flux by the direct integration method \cite{atkins03}.  Due to the fact that the direct integration needs stable performance from the detector, the lifetime of the the analysis is reduced to 290 days. Finally, since the integration time used is 24 hours due to the size of the\textit{Fermi} Bubbles, effects from the large-scale anisotropy needs to be removed \cite{aniso14}. The analysis is done in 7 analysis bins defined in the similar way as in \cite{crabpaper}. 

After taking into account these effects, the excess calculation is given by
\begin{equation}
G'_i = \varepsilon_{G,i} \frac{E'_i-\varepsilon_{C,i}E_i}{\varepsilon_{G,i}-\varepsilon_{C,i}}, 
\end{equation}
where, $G'_i$ is the final excess after corrections in pixel $i$; $E'_i$ is the excess in pixel $i$ without corrections after applying gamma-hadron cuts; $E_i$ is the excess in pixel $i$ without corrections before applying gamma-hadron cuts; $\varepsilon_{G,i}$ and $\varepsilon_{C,i}$ are the gamma and hadron efficiencies after applying gamma-hadron cuts.
For more details on the analysis see \cite{hawcbubble}.

\subsection{Upper Limits}
\begin{figure}
\centering
\includegraphics[scale=0.4]{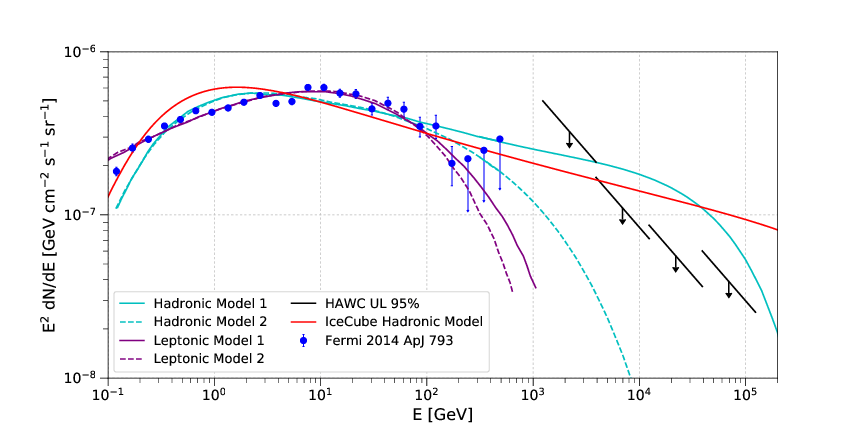}
\caption{Measured flux of the Fermi Bubbles with HAWC upper limits. Hadronic and leptonic models that explain the emission of the Bubbles are overlaid too. }
\label{fig:bubble}
\end{figure}

No significant excess was found, we proceeded to calculate upper limits on the flux. Figure \ref{fig:bubble},  shows the data points from the \textit{Fermi} measurements, together with the HAWC upper limits. The energy bins are obtained by combining the analysis bins with a weighted average (see \cite{hawcbubble} for the description).
The plot also features predictions from two hadronic and two leptonic models, all obtained from \cite{ackerman14}. 
The leptonic models are obtained from an electron spectrum with the shape of a power-law with exponential cutoff interacting with the interstellar radiation field at 5\,kpc from the Galactic plane and the cosmic microwave background. 
The hadronic models assume a power-law and power-law with a cutoff for the proton spectrum. The protons interact with the interstellar medium and produce photons through pion decay.  
The IceCube model is obtained from \cite{lunardini15} and it is the counterpart of a neutrino flux model that best fist the IceCube data.
Our result is not able to constrain the models at energies below 1\,TeV. However at high energies, it implies, for a hadronic model, that there is a cutoff in the proton spectrum. 

\section{Conclusion}

Observations of large gamma-ray structures can give us an insight in how cosmic rays propagate and distribute in the galaxy, as well as the mechanisms that produce them. This information is useful to understand the evolution of our galaxy. 

Using data from the HAWC observatory, we searched for gamma-ray signal in three GMCs and the Fermi Bubbles. In the case of the Fermi Bubbles we calculated upper limits at $95\%$ C.L. The upper limits constrain some hadronic models, including a neutrino model that describes the IceCube data.

\acknowledgments{
We	acknowledge	the	support	from:	the	US	National	Science	Foundation	(NSF);	the	
US	Department	of	Energy	Office	of	High-Energy	Physics;	the	Laboratory	Directed	
Research	and	Development	(LDRD)	program	of	Los	Alamos	National	Laboratory;	
Consejo	Nacional	de	Ciencia	y	Tecnolog\'{\i}a	(CONACyT),	M{\'e}xico	(grants	
271051,	232656,	260378,	179588,	239762,	254964,	271737,	258865,	243290,	
132197),	Laboratorio	Nacional	HAWC	de	rayos	gamma;	L'OREAL	Fellowship	for	
Women	in	Science	2014;	Red	HAWC,	M{\'e}xico;	DGAPA-UNAM	(grants	RG100414,	
IN111315,	IN111716-3,	IA102715,	109916,	IA102917);	VIEP-BUAP;	PIFI	2012,	
2013,	PROFOCIE	2014,	2015; the	University	of	Wisconsin	Alumni	Research	
Foundation;	the	Institute	of	Geophysics,	Planetary	Physics,	and	Signatures	at	Los	
Alamos	National	Laboratory;	Polish	Science	Centre	grant	DEC-2014/13/B/ST9/945;	
Coordinaci{\'o}n	de	la	Investigaci{\'o}n	Cient\'{\i}fica	de	la	Universidad	
Michoacana. Thanks	to	Luciano	D\'{\i}az	and	Eduardo	Murrieta	for	technical	
support.
}

\end{document}